\PassOptionsToPackage{font=small}{caption}
\documentclass[a4paper]{jpconf}
\pdfoutput=1
\usepackage[T1]{fontenc}
\usepackage{setspace}
\usepackage{tabulary}
\usepackage{subfig}
\usepackage{graphicx}
\usepackage{amsmath}
\usepackage{adpwrapfig}
\usepackage{multirow}
\usepackage{iopams}
\begin{document}
%\vspace{-80mm}
\title{Using MiniBooNE NCEL and CCQE cross section results to constrain 3+1 sterile neutrino models}
\author{C Wilkinson, S Cartwright and L Thompson}
\address{Department of Physics and Astronomy, University of Sheffield,\\Hicks Building, Hounsfield Road, Sheffield, S3 7RH, United Kingdom}

\ead{callum.wilkinson@sheffield.ac.uk}
%\vspace{-2mm}
\begin{abstract} The MiniBooNE NCEL and CCQE cross-section measurements (neutrino running) are used to set limits in the $\Delta m^{2}-\sin^{2}\vartheta_{\mu s}$ plane for a 3+1 sterile neutrino model with a mass splitting $0.1 \leq \Delta m^{2} \leq 10.0$ eV$^{2}$. GENIE is used, with a relativistic Fermi gas model, to relate $E_{\nu}$ and the reconstructed quantities measured. The issue of uncertainty in the underlying cross-section model and its effect on the sterile neutrino limits is explored, and robust sterile neutrino limits are produced by fitting the sterile parameters and the axial-mass cross-section parameter simultaneously.
\end{abstract}
%\vspace{-12mm}
\section{Introduction}
\label{sec:intro}The large axial-mass ($M_{A}$) measured by MiniBooNE and other experiments has shown that simple RFG models are inadequate to describe experimental data from quasi-elastic neutrino scattering off nuclear targets. Although there has been a great deal of recent theoretical work developing more sophisticated cross-section models, a clear picture has yet to emerge (a recent summary can be found in~\cite{zeller12}). Neutrino oscillation experiments use the measured event rate to infer detailed information about the flux, so a flawed cross-section model may bias results. There have been a number of studies investigating this bias in the context of three-neutrino mixing measurements (see for example ~\cite{huberBias13, moselBias13, benharBias13}). Similarly, such a bias should be investigated for sterile neutrino results, which may be more susceptible as there is generally no way to measure the unoscillated flux.

This work investigates the effect that uncertainty in an RFG cross-section model, with $M_{A}$ as the only free parameter, has on sterile limits produced by a simple analysis of MiniBooNE NCEL and CCQE cross-section data. It extends the work published in~\cite{mbNCELFit13}, which omitted the CCQE data because of the lack of bin correlations. Limits are set in the $\Delta m^{2}-\sin^{2}\vartheta_{\mu s}$ plane for a 3+1 sterile neutrino model with a mass splitting $0.1 \leq \Delta m^{2} \leq 10.0$ eV$^{2}$ using a number of different assumptions about the RFG model. We implicitly follow the assertion made in~\cite{mbAntiNCEL}, that inflating $M_{A}$ provides a reasonable description of the data, though it is understood that this inflated $M_{A}$ value is effectively accounting for additional nuclear effects. We will refer to the inflated axial-mass as $M_{A}^{\mbox{\scriptsize{eff}}}$ from now on. A worthwhile extension of this work would be to look at the effect that different cross-section models have on the sterile neutrino limits produced.

There are two choices to be made regarding the simple cross-section model, and we demonstrate that their effects on the sterile limits are significant. The first is whether to sequentially fit $M_{A}^{\mbox{\scriptsize{eff}}}$ then the sterile neutrino parameters, which is only statistically sound if $M_{A}^{\mbox{\scriptsize{eff}}}$ and the sterile parameters are completely uncorrelated, or fit all parameters simultaneously. The former procedure was used in the MiniBooNE-SciBooNE sterile analyses~\cite{mbSciNu2012, mbSciANu2012}, which used the MiniBooNE measurement of $M_{A}^{\mbox{\scriptsize{eff}}}$ as a constrained parameter in the fit, though it was noted that the cross-section and sterile parameters had been found to be nearly uncorrelated. In general, however, it may not be obvious that an experiment which has a prior {\it in situ} measurement of their cross-section parameters must investigate correlations with the sterile parameters. The second choice is whether to fit a separate $M_{A}^{\mbox{\scriptsize{eff}}}$ value for the NCEL and CCQE samples, or whether one value should be used. The correct choice is not clear without a full understanding of the nuclear effects being covered by $M_{A}^{\mbox{\scriptsize{eff}}}$.

It should be noted that constraints from other experiments cannot be used because the nuclear effects being modelled by $M_{A}^{\mbox{\scriptsize{eff}}}$ are detector and neutrino energy dependent. Of course, when more sophisticated models emerge which provide a consistent description of all the available experimental data, these issues will be resolved.
\section{Analysis method}
\begin{wraptable}{R}{0.6\textwidth}
%\vspace{-12mm}
\footnotesize
  \centering
    {\renewcommand{\arraystretch}{1.2}
    \begin{tabulary}{0.55\textwidth}{CCCC}
	  \hline\hline
	  {\bf Property} & {\bf MiniBooNE NCEL} & {\bf MiniBooNE CCQE} \\\hline\hline
	  Baseline L (m) & 541 & 541 \\
	  Average Neutrino Energy (GeV) & 0.788 & 0.788 \\
	  Energy Range for Measurement (GeV) & $0 \leq E_{\nu} \leq 10$ & $0 \leq E_{\nu} \leq 3$\\
	  Signal Events & $\nu_{\mu, e} + n,p \rightarrow \nu_{\mu, e} + n,p$ & $\nu_{\mu} + n \rightarrow \mu^{-} + p$\\
	  POT & $6.46 \times 10^{20}$ & $5.58 \times 10^{20}$\\
	  Integrated Flux $\Phi_{\nu}$ ($\nu$ cm$^{-2}$ POT$^{-1}$) & $5.22227 \times 10^{-10}$ & $5.16 \times 10 ^{-10}$\\
	  Target Material & CH$_{2}$ & CH$_{2}$ \\
	  \hline\hline	
      \end{tabulary}}
      \caption{Summary of the important experimental details for the two samples used in this analysis. Further details describing the NCEL sample can be found in~\cite{mbNCEL, PerevalovThesis}, and for the CCQE sample in~\cite{mbCCQE, KatoriThesis}.}	
  \label{tab:signal}
%\vspace{-6mm}
\end{wraptable}
Relating reconstructed quantities (the published cross-section results) with $E_{\nu}$ requires a cross-section model. Here we make this model dependence explicit by using GENIE to simulate events on the MiniBooNE detector material, CH$_{2}$, with our chosen RFG model, and then producing a flux-averaged cross-section prediction to compare with the published results.

A full description of the method used to produce model predictions can be found in~\cite{mbNCELFit13} for the NCEL dataset. The extension to include CCQE is straightforward, using the same cross-section model with the signal definition given in Table~\ref{tab:signal}. The MiniBooNE flux prediction is given in~\cite{mbFlux}.
%\vspace{-10pt}
\section{Fit Details}
We follow the fit procedure described in~\cite{mbNCELFit13}, with a modified $\chi^{2}$ expression, given in Equation~\ref{eq:chiSq}, which includes the additional CCQE bins. 
\footnotesize
\begin{align}
\indent
\notag\chi^{2} (\mathbf{\theta}) &= \left \lbrack \sum^{51}_{i=0}\sum^{51}_{j=0} \left(\nu_{i}^{DATA} - \nu_{i}^{MC}(\mathbf{\theta}) \right) M_{ij}^{-1} \left(\nu_{j}^{DATA} - \nu_{j}^{MC}(\mathbf{\theta}) \right) + \left(\frac{\mathbf{\theta}_{M_{A}}}{\sigma_{M_{A}}} \right)^{2} \right \rbrack \rightarrow \chi^{2}_{NCEL}(\theta)\\
&+ \left \lbrack \sum_{k=0}^{17} \left ( \frac{\nu_k^{DATA} - \xi\nu_{k}^{MC}(\mathbf{\theta})}{\sigma_{k}} \right )^{2} + \left( \frac{\xi - 1}{\sigma_{\xi}} \right)^{2}\right \rbrack \rightarrow \chi^{2}_{CCQE}(\theta),
\label{eq:chiSq}
\end{align}
\normalsize
\noindent where $\xi$ is the MiniBooNE normalisation factor, varied in all fits, and $\sigma_{\xi}$ is the published CCQE normalisation uncertainty of 10.7\%~\cite{mbCCQE}.

Note that for sequential fits, additional penalty terms are added to the $\chi^{2}$ from Equation~\ref{eq:chiSq} for each $M_{A}^{\mbox{\scriptsize{eff}}}$ parameter included (a full description can be found in~\cite{mbNCELFit13}). The error and central values for these penalty terms are taken from fits where $M_{A}^{\mbox{\scriptsize{eff}}}$ is varied, and no sterile mixing is assumed, the results of which are given in Table~\ref{tab:maOnlyJoint}.
\section{Results}
\begin{figure}[h!]
% \vspace{-20pt}
 \centering
  \subfloat[$M_{A}^{\mbox{\scriptsize{BOTH}}}$ sequential]                     {\includegraphics[width=0.4\textwidth]{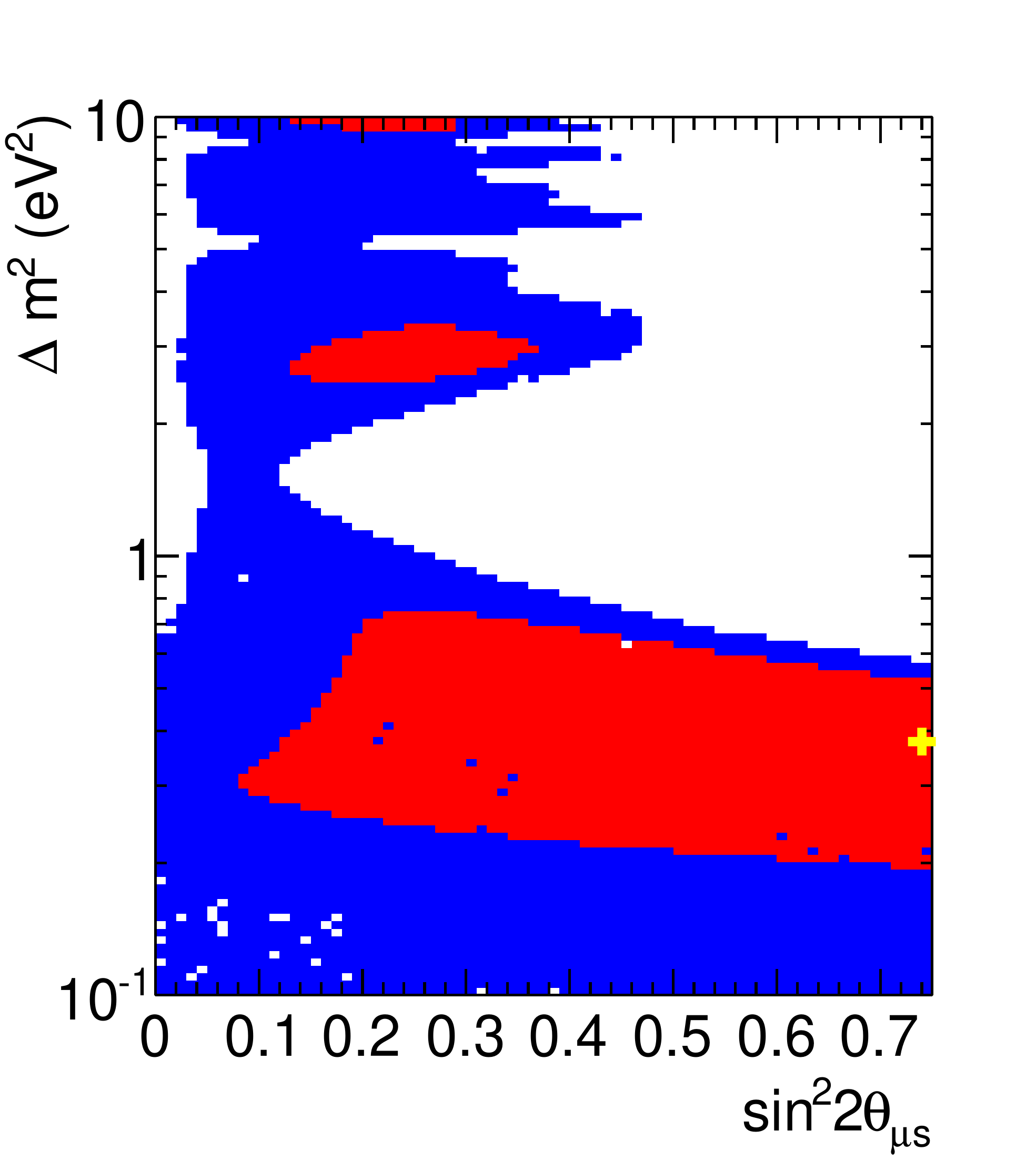} \label{subfig:pScan_SeqTog}}
  \subfloat[$M_{A}^{\mbox{\scriptsize{CCQE}}}$ \& $M_{A}^{\mbox{\scriptsize{NCEL}}}$ sequential]   {\includegraphics[width=0.4\textwidth]{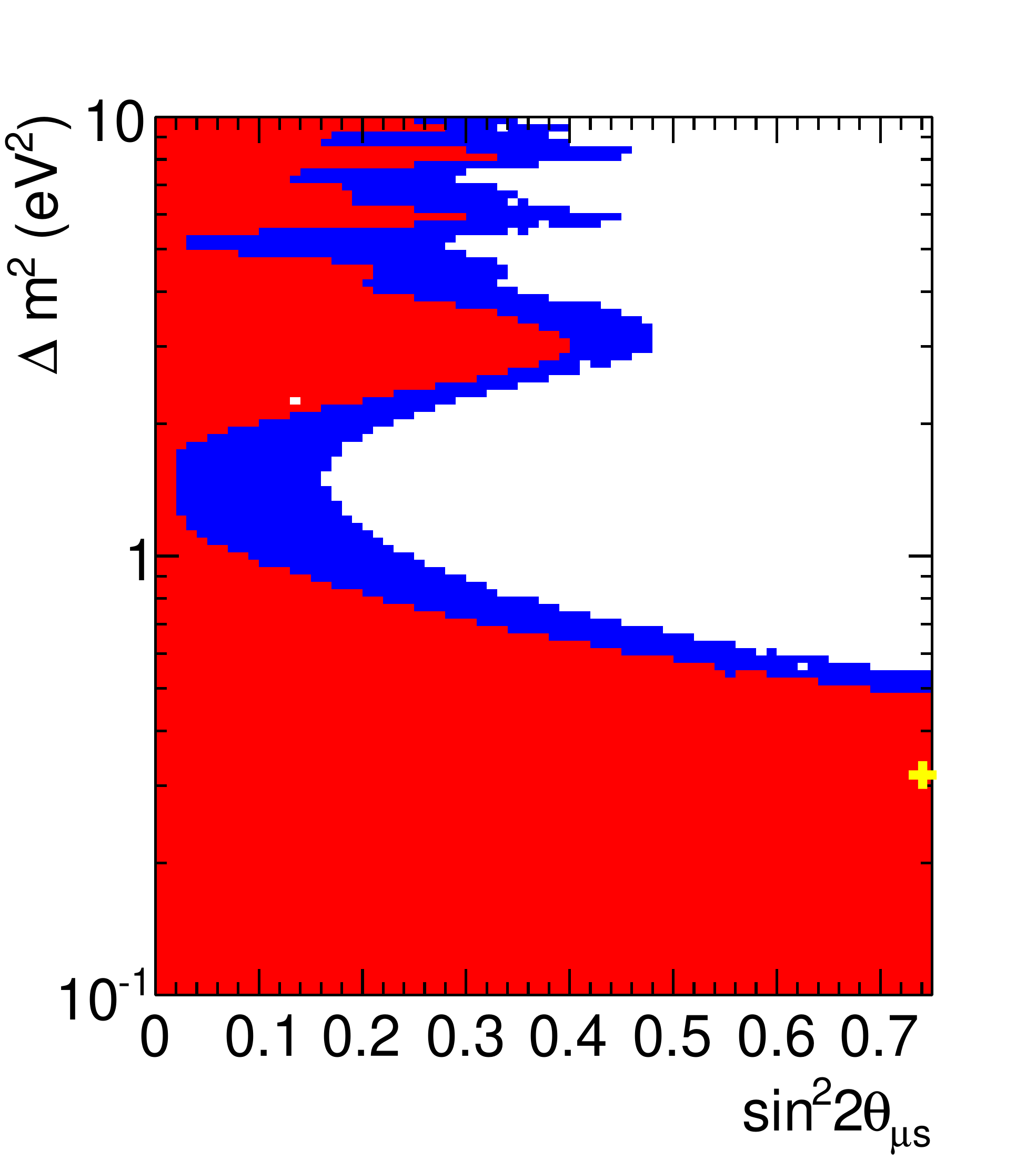} \label{subfig:pScan_SeqSep}}\\
  \subfloat[$M_{A}^{\mbox{\scriptsize{BOTH}}}$ simultaneous]                   {\includegraphics[width=0.4\textwidth]{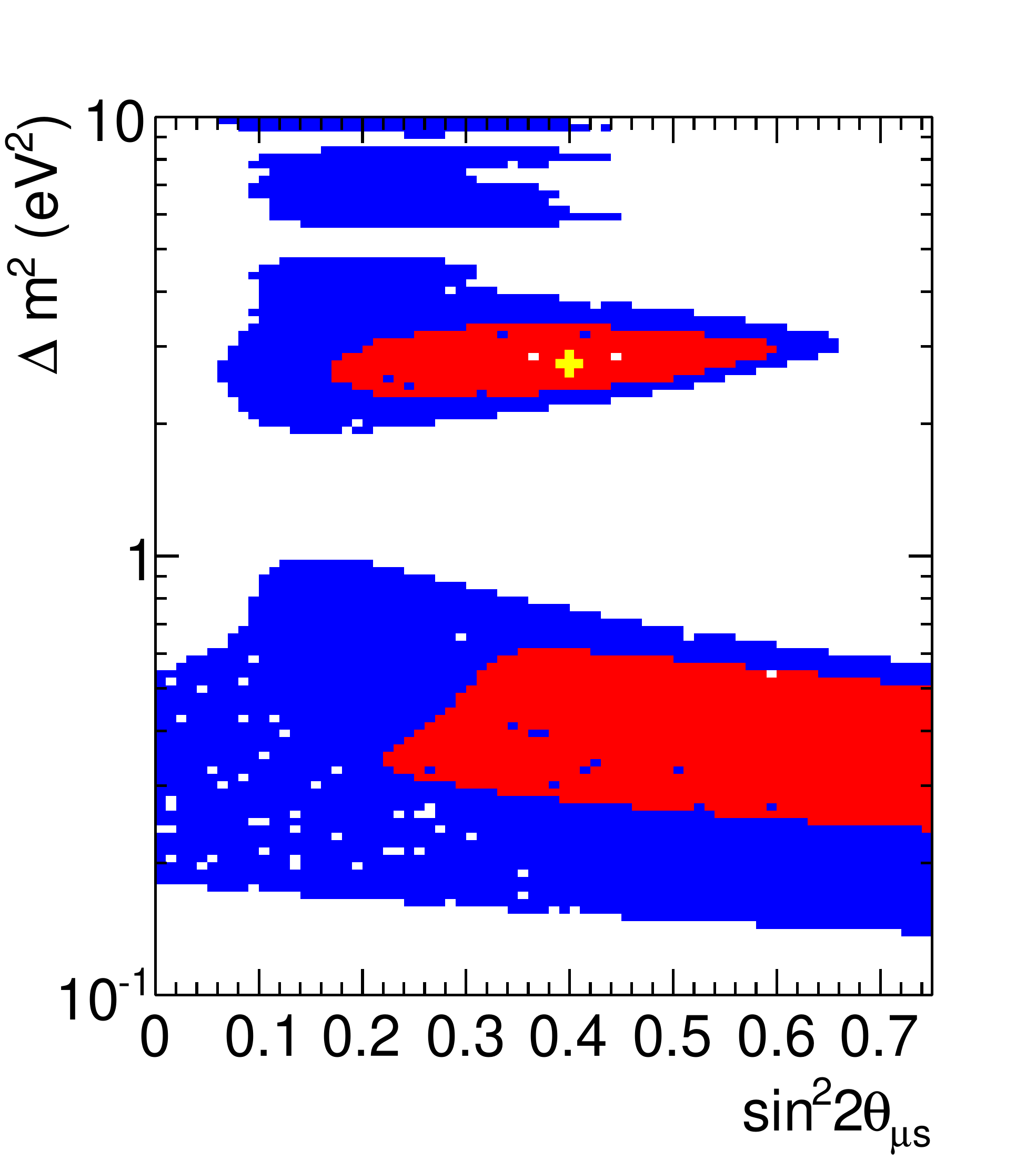} \label{subfig:pScan_SimTog}}
  \subfloat[$M_{A}^{\mbox{\scriptsize{CCQE}}}$ \& $M_{A}^{\mbox{\scriptsize{NCEL}}}$ simultaneous] {\includegraphics[width=0.4\textwidth]{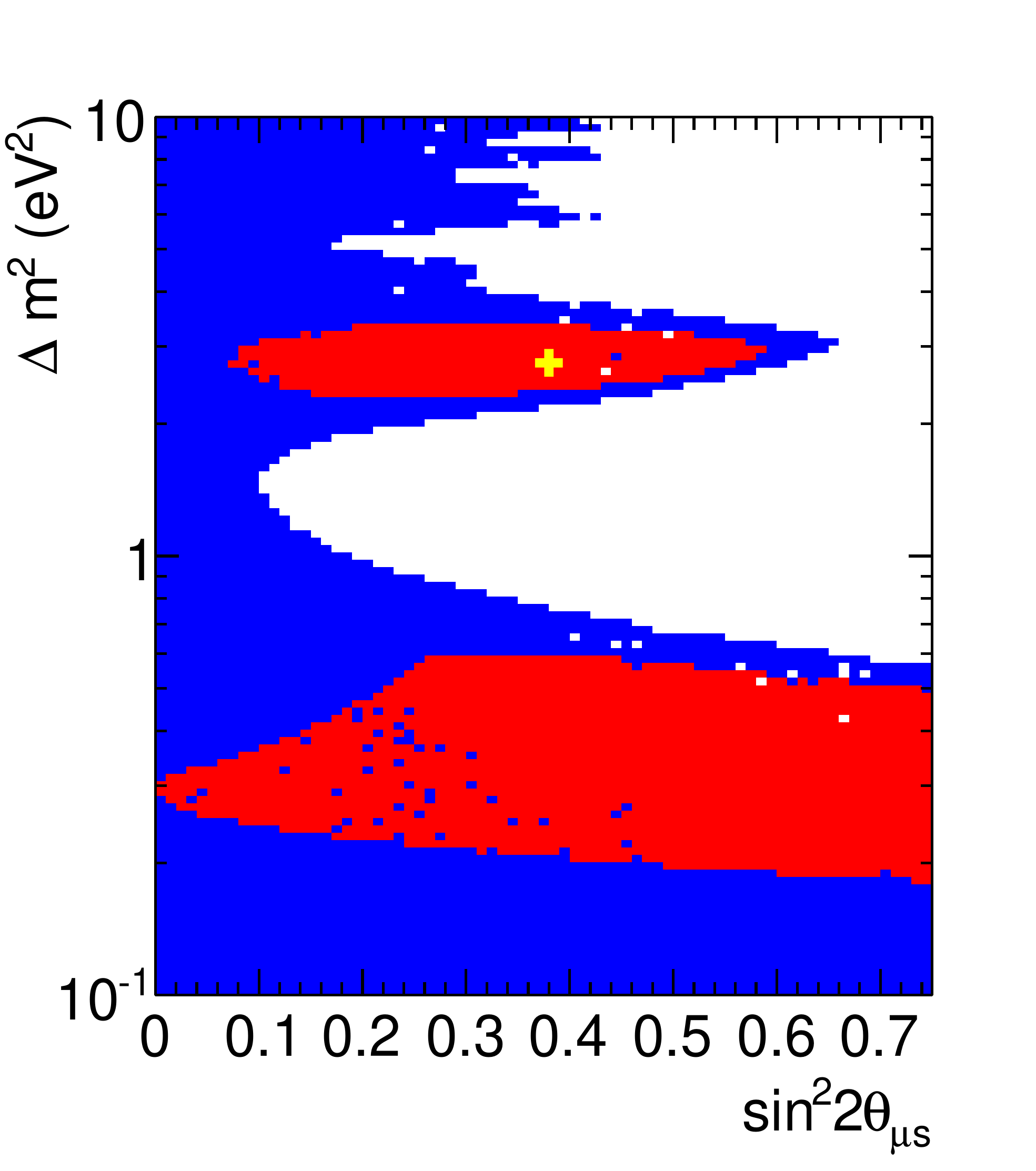} \label{subfig:pScan_SimSep}}
% \vspace{-5pt}    
  \caption{The exclusion plots produced by joint fits to both NCEL and CCQE datasets. The 90\% region is shown in red, the 99\% region is shown in blue, and the best fit point is indicated with a yellow cross.}\label{fig:pScansJoint}
% \vspace{-5pt}
\end{figure}
Joint fits to both datasets are shown in Figure~\ref{fig:pScansJoint}, sequential and simultaneous fits are shown, both with a common $M_{A}^{\mbox{\scriptsize{BOTH}}}$ value for both NCEL and CCQE datasets, and with separate $M_{A}^{\mbox{\scriptsize{NCEL}}}$ and $M_A^{\mbox{\scriptsize{CCQE}}}$ values. The best fit $\chi^{2}$ and parameter values are given in Table~\ref{tab:bestFitsJoint}. 
\section{Analysis and Conclusions}
%\begin{wrapfigure}{L}{0.6\textwidth}
\begin{figure}[htb]
  \centering
%  \vspace{-15pt}
  \includegraphics[width=0.4\textwidth]{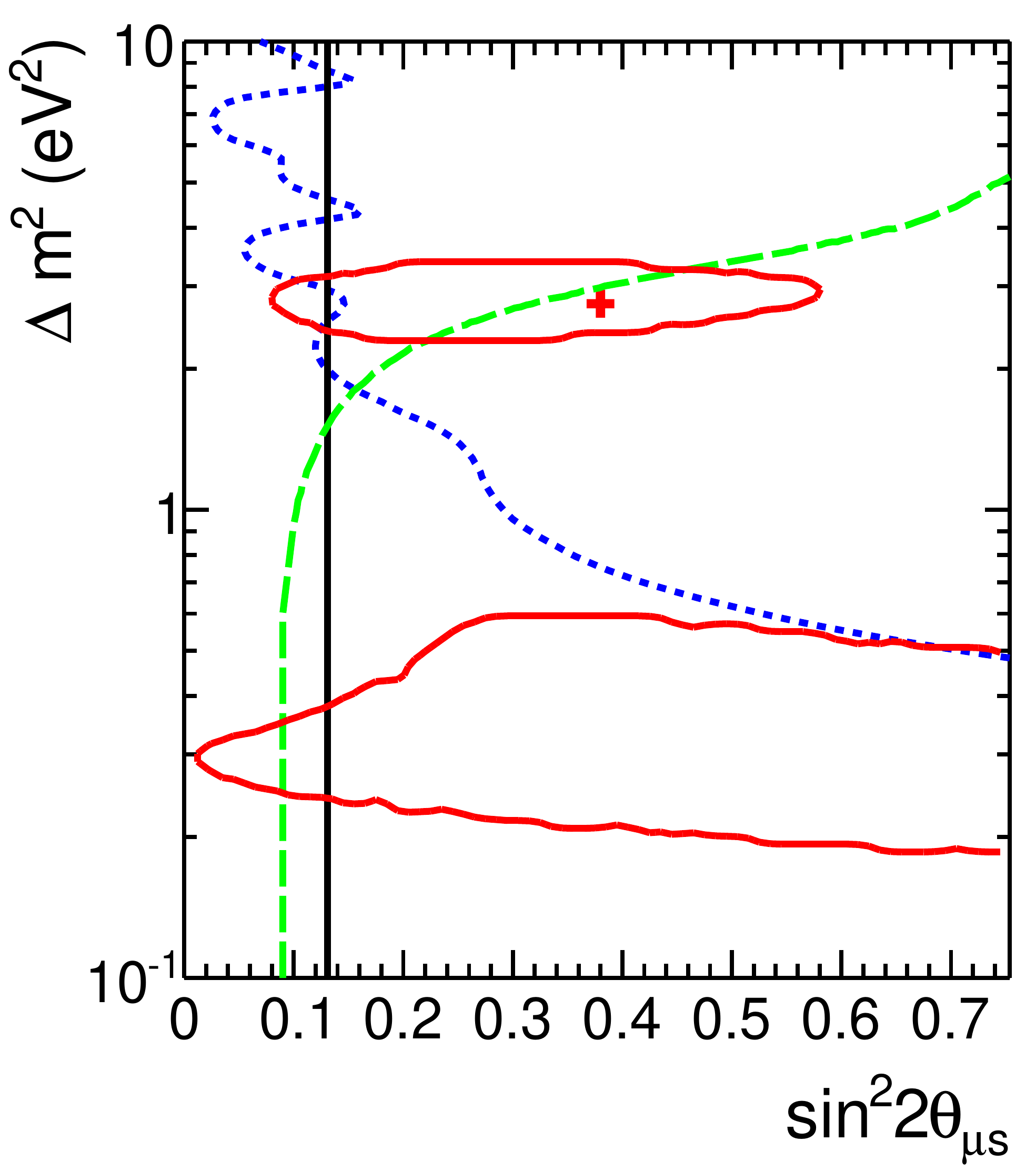}
%  \vspace{-10pt}
  \scriptsize
   \caption{90\% confidence region from the simultaneous fit (solid red line with best fit red cross), with MiniBooNE-SciBooNE $\nu_{\mu}$-disappearance limits~\cite{mbSciNu2012} (short dashed blue line), atmospheric limits~\cite{maltoni07} (black solid line), and MINOS NC-disappearance analysis~\cite{giunti11MINOS} (long dashed green line).}
   \label{fig:sterileComparison}
%\end{wrapfigure}
\end{figure}
It is clear from Figure~\ref{fig:pScansJoint} that there are differences between sequential and simultaneous fitting, indicating that the sterile and cross-section model parameters are correlated in this case, and therefore the sequential fit is not reliable. We should note that the correlation is strong in the NCEL case, and weak in the CCQE case, so the assertion that they are uncorrelated in~\cite{mbSciNu2012, mbSciANu2012} is probably reasonable, but in general this cannot be assumed.

There is also a considerable difference between the contours produced when fitting to one or two $M_{A}^{\mbox{\scriptsize{eff}}}$ terms, as can be seen by comparing Figures~\ref{subfig:pScan_SimTog} and ~\ref{subfig:pScan_SimSep}. The correct choice is not clear, so in Figure~\ref{fig:sterileComparison} we take the more conservative limits, with separate $M_A^{\mbox{\scriptsize{CCQE}}}$ and $M_{A}^{\mbox{\scriptsize{NCEL}}}$ parameters, and compare the 90\% confidence regions with existing datasets. It is clear that there is strong disagreement with other limits. Note that in Figure~\ref{fig:sterileComparison} we use the relation $\sin^{2}2\theta_{\mu s} \leq \sin^{2}2\theta_{\mu\mu}$ to plot other results in the same plane. We interpret the disagreement with other sterile neutrino results as evidence that the cross-section model choice has resulted in tension between the 
NCEL and CCQE datasets, which was resolved by favouring more sterile mixing. However, it should also be noted that this result is perfectly consistent with the methods used to produce other sterile mixing results, and could also be interpreted as evidence that adding the NCEL dataset gives additional power to constrain the 3+1 model in this plane. The tension with other datasets can be interpreted as evidence that the 3+1 model is insufficient to describe all of the sterile neutrino data available. 
\begin{table}[!h]
\footnotesize
  \centering
  \parbox{.45\textwidth}{
    {\renewcommand{\arraystretch}{1.2}  
    \begin{tabulary}{0.45\textwidth}{CCCC}
	    \hline\hline
	   & {\bf Fit} & {\bf $\chi^{2}$}/DOF & {\bf M$_{A}$} \\\hline\hline
	  \multirow{3}{*}{This analysis} & NCEL & 32.1/50 & 1.24 $\pm$ 0.08 \\
	  & CCQE & 20.2/16 & 1.46 $\pm$ 0.05 \\
	  & Joint & 57.0/67 & 1.40 $\pm$ 0.04 \\\hline
	  \multirow{2}{*}{MiniBooNE} & NCEL~\cite{mbNCEL} & 26.9/50 & 1.39 $\pm$ 0.11 \\
	  & CCQE~\cite{mbCCQE} & ---/17 & 1.35 $\pm$ 0.17 \\
	  \hline\hline	
      \end{tabulary}}
  \caption{Best fit values for each of the $M_{A}^{\mbox{\scriptsize{eff}}}$ fits performed for this analysis, with published MiniBooNE values for comparison.}	
  \label{tab:maOnlyJoint}}%      
  \hfill
  \parbox{.5\textwidth}{
      {\renewcommand{\arraystretch}{1.2}
      \begin{tabulary}{0.5\textwidth}{CCCCC}
	  \hline\hline
	  \multirow{2}{*}{Fit Description} & \multicolumn{2}{c}{Sequential} & \multicolumn{2}{c}{Simultaneous} \\
	  & TWO & ONE & TWO & ONE \\\hline\hline
	  $\chi^{2}$/DOF              & 47.3/47  & 46.8/47               & 44.1/45  & 44.6/46 \\
	  $\Delta m^{2}$              & 0.32  & 0.38               & 2.75  & 2.74  \\
	  U$_{e4}$                    & 5.1$\times 10^{-2}$  & 4.8$\times 10^{-7}$ & 3.9$\times 10^{-2}$ & 2.7$\times 10^{-7}$ \\
	  U$_{\mu4}$                  & 0.50  & 0.46                & 0.33  & 0.34  \\
	  sin$^{2}2\vartheta_{\mu s}$ & 0.74  & 0.74                & 0.38  & 0.40  \\
	  M$_{A}^{\mbox{\scriptsize{NCEL}}}$                & 1.26  & 1.38                  & 1.52  &  1.62   \\
	  M$_{A}^{\mbox{\scriptsize{CCQE}}}$                & 1.43  & 1.38                  & 1.62  &  1.62   \\
	  CCQE Norm.                  & 1.10   & 1.16                & 1.24  & 1.26  \\
          \hline\hline	
      \end{tabulary}}
  \caption{Best fit values for all of the fits performed. Each fit uses either one or two $M_{A}^{\mbox{\scriptsize{eff}}}$ parameters. The sequential fits use the relevant $M_{A}^{\mbox{\scriptsize{eff}}}$ values and errors calculated in Table~\ref{tab:maOnlyJoint} in the penalty terms.}	
  \label{tab:bestFitsJoint}}
\end{table}
\newline
%\begin{spacing}{0.9}
\bibliographystyle{iopart-num}
\bibliography{steriles}
%\end{spacing}
\end{document}